\documentclass[prr,amsmath, amssymb, aps, floatfix, superscriptaddress, twocolumn, noeprint]{revtex4-2}
\usepackage{amsmath}
\usepackage{graphicx}
\usepackage{dcolumn}
\usepackage{bm}
\usepackage{xcolor}
\usepackage{hyperref}
\hypersetup{
    colorlinks=true,
    linkcolor=blue,
    citecolor=blue,
    filecolor=magenta,
    urlcolor=blue,
    breaklinks=true
} 
\usepackage{verbatim}
\usepackage{braket}
\usepackage{siunitx}
\usepackage{makecell}
\usepackage{lineno}
\usepackage{float}
\usepackage{multirow}

\usepackage{notes2bib}

\usepackage[linesnumbered,ruled,vlined]{algorithm2e}

\begin{document}

\title{Diagnosing Quantum Many-body Chaos in Non-Hermitian Quantum Spin Chain via Krylov Complexity}
\author{Yijia Zhou}
\affiliation{Shanghai Qi Zhi Institute, Shanghai 200232, China }
\author{Wei Xia}
\affiliation{Department of Physics, The Chinese University of Hong Kong, Shatin, New Territories, Hong Kong, China}
\author{Lin Li}
\affiliation{National Gravitation Laboratory, MOE Key Laboratory of Fundamental Physical Quantities Measurement, Hubei Key Laboratory of Gravitation and Quantum Physics, Institute for Quantum Science and Engineering, School of Physics, Huazhong University of Science and Technology, Wuhan 430074, China}
\author{Weibin Li}
\affiliation{School of Physics and Astronomy, and Centre for the Mathematics and Theoretical Physics of Quantum Non-equilibrium Systems, The University of Nottingham, Nottingham NG7 2RD, United Kingdom}

\begin{abstract}
We investigate the phase transitions from chaotic to nonchaotic dynamics in a quantum spin chain with a local non-Hermitian disorder, which can be realized with a Rydberg atom array setting. As the disorder strength increases, the emergence of nonchaotic dynamics is qualitatively captured through the suppressed growth of Krylov complexity, and quantitatively identified through the reciprocity breaking of Krylov space. We further find that the localization in Krylov space generates another transition in the weak disorder regime, suggesting a weak ergodicity breaking. Our results closely align with conventional methods, such as the entanglement entropy and complex level spacing statistics, and pave the way to explore non-Hermitian phase transitions using Krylov complexity and associated metrics.
\end{abstract}

\date{\today}
\maketitle

\section{Introduction}

Diagnosis of chaos emerging in quantum systems attracts a growing interest, with a variety of measures including Lyapunov exponents \cite{Cerdeira1988, Chavez-Carlos2019, Garcia-Garcia2024}, out-of-time-order correlations~\cite{Garttner2017, Chen2017, Xu2024}, Loschmidt echo~\cite{Gorin2006, Yan2020a, Hasegawa2021}, and spectral characteristics revealed through level spacing distributions~\cite{Atas2013, Srivastava2019, Sa2020, Hamazaki2020, Garcia-Garcia2022, Kawabata2023a, Roccati2024} and spectral form factors \cite{Berry1985, Bertini2018, Dong2025}.
Recently, Krylov complexity has emerged as an innovative approach to the study of quantum chaos~\cite{Parker2019, Dymarsky2020, Balasubramanian2022, BallarTrigueros2022, Rabinovici2022, Rabinovici2022a, Erdmenger2023, Hashimoto2023, Espanol2023, Bhattacharyya2023, Bhattacharyya2023a, Gautam2024, Baggioli2025, Scialchi2024, Menzler2024, Camargo2024a, Chen2025, Bhattacharjee2025, Li2025, Nandy2025a, Nandy2025c, Nandy2025}. Krylov complexity quantifies the spreading of quantum operators \cite{Parker2019} or states~\cite{Balasubramanian2022} in Krylov space, capturing key features of the underlying dynamics. 

\begin{figure}
    \centering
    \includegraphics[width=0.97\linewidth]{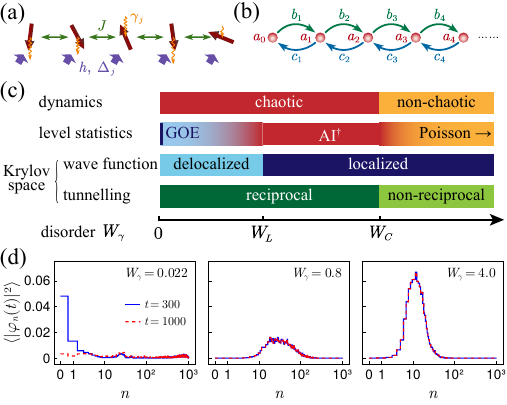}
    \caption{ 
    (a) Disordered non-Hermitian XY model, with XY coupling $J$, transverse field $h$, site-dependent detuning $\Delta_j \in  [-W_\Delta, W_\Delta]$, and dissipation $\gamma_j \in [-W_\gamma, W_\gamma]$.
    (b) Illustration of the Krylov chain, which maps the physical system to the Krylov space. The Krylov wave function is initially localized at the first site and then propagate throughout the chain. Tunnelling rates $b_j$ and $c_j$ are generally not conjugate for non-Hermitian systems. 
    (c) Sketch of the phase diagram with increasing $W_{\gamma}$. The dynamics exhibits a chaos to nonchaos phase transition as the disorder term $W_\gamma$ increases. At $W_\gamma = 0$, the system follows Gaussian orthogonal ensemble (GOE) statistics, transitioning to the AI$^\dagger$ class near $W_L$ and then approaching Poisson statistics for very large $W_{\gamma}$. In Krylov space, wave functions remain delocalized for $W_\gamma < W_L$ at long times, but localized for $W_\gamma > W_L$. The first few tunneling coefficients, $b_j$ and $c_j$, show a transition from reciprocal to nonreciprocal regimes at $W_C$, which enhances localization.
    (d) Krylov wave functions in chaotic regimes with delocalized (left) and localized (middle) profiles, and nonchaotic (right) regimes, at mid-term $t=300$ (solid blue) and long-term $t=1000$ (dashed red). See text for details.}
    \label{fig:model}
\end{figure}

A pivotal insight in this field is the operator growth hypothesis~\cite{Parker2019}, which suggests that information scrambling and chaos can be identified by observing exponential growth in Krylov complexity at early times. The exponential growth is attributed to the linearity of Lanczos coefficients, which govern the tunneling rates in Krylov space. 
In contrast, another hypothesis focuses on the long-term dynamics and proposes a Tsallis $q$-log behavior of the Lanczos coefficients throughout the entire Krylov space~\cite{Erdmenger2023, Bhattacharjee2025, Nandy2025}, where chaos can be signified by the linear growth in complexity at intermediate times, followed by a peak before saturation~\cite{Balasubramanian2022, Erdmenger2023, Baggioli2025}. Additionally, the fluctuation in Lanczos coefficients, known as Krylov variance, has been shown to correlate with the delocalization of the Krylov wave function and, consequently, with ergodic behaviors~\cite{Dymarsky2020, BallarTrigueros2022, Hashimoto2023, Menzler2024, Scialchi2024, Li2025}. Despite extensive investigations in the Hermitian regime, challenges emerge when extending to open quantum systems. While the complexity appears to be significantly reduced under dissipation~\cite{Bhattacharya2022, Bhattacharjee2023, Liu2023, Bhattacharya2023, Bhattacharjee2024, Srivatsa2024, Carolan2024, Bhattacharya2024}, the Lanczos coefficients deviate from the established hypotheses for Hermitian systems, particularly when the absence of Hermiticity of Krylov space is overlooked. Furthermore, the critical behavior and universal properties near phase transitions in open systems remain largely unexplored.

In this work, we investigate Krylov complexity of a non-Hermitian XY model with random local gain and loss. We show that the evolution of Krylov wave functions and complexity growth are indicative of phase transitions of quantum chaos. 
Through a closer examination of Lanczos coefficients, we find that the Krylov variance elucidates critical points and universal features for localization of the wave function. This indicates a weak ergodicity breaking as the quantum state can only evolve within a confined Krylov subspace. Additionally, in the delocalization regime, the Krylov complexity growth behavior exhibits a prethermal plateau before saturation, which links to the Hermiticity breaking~\cite{Akemann2025}. To identify transitions between chaotic and nonchaotic phases, we introduce a metric of reciprocity, where the onset of nonreciprocity of Krylov space marks the corresponding critical point.
Our results are consistent with the complex level spacing statistics~\cite{Sa2020, Hamazaki2020, Garcia-Garcia2022}, which reflect the conjectures of level repulsion in quantum chaos by Bohigas, Giannoni, and Schmit for closed systems~\cite{Bohigas1984} and Grobe, Haake, and Sommers for open systems~\cite{Grobe1988}.
We have further compared the critical points using our method with the variance of entanglement entropy~\cite{Kjall2014, Hamazaki2019}, which are also qualitatively consistent.

Our framework is applicable to other ergodicity-breaking phenomena, including many-body localization~\cite{BallarTrigueros2022, Ganguli2024a, DeTomasi2024}, quantum scars~\cite{Bhattacharjee2022, Nandy2024}, and Hilbert space fragmentation~\cite{Balasubramanian2024, Desaules2024}, and holds potential for extensions into machine learning applications~\cite{Cindrak2024}.
The non-Hermitian XY model can be realized with Rydberg atom array quantum simulators~\cite{Weimer2010, Wu2021, Morgado2021} and circuit-based quantum simulators~\cite{Suzuki2022, Suri2023, Ghanem2023, Suzuki2025}. Our work thus opens avenues to experimentally explore complexities with observables such as Loschmidt echo~\cite{Gorin2006, Chenu2019, Porter2022, Tu2023, Geier2024}. 

This paper is organized as follows. In Sec.~\ref{sec:model}, we describe the non-Hermitian XY model and briefly show the phase transitions about quantum chaos. In Sec.~\ref{sec:Krylov}, we present the concept and definition of Krylov complexity and relevant quantities, and show the overall features of complexity growth under different parameters. In Sec.~\ref{sec:results}, we use the Krylov variance and reciprocity to quantitatively identify the critical behaviors of phase transitions. Finally, in Sec.~\ref{sec:benchmark}, we compare our results with the conventional signatures including the complex level spacing (CSR) ratio and entanglement entropy.

\section{Model}
\label{sec:model}

We study a one-dimensional disordered XY spin chain under both transverse and longitudinal fields, as depicted in Fig.~\ref{fig:model}(a). The Hamiltonian is given by
\begin{equation}
H=\sum_{j=1}^{L-1} J \left( {\hat{\sigma}_{j}^{x}\hat{\sigma}_{j+1}^{x}+\hat{\sigma}_{j}^{y}\hat{\sigma}_{j+1}^{y}} \right) +\sum_{j=1}^{L} {h\hat{\sigma}_{j}^{x}+\mathcal{D}_j\hat{\sigma}_{j}^{z}}, \label{eq:Hamiltonian}
\end{equation}
where $\hat{\sigma}_{j}^{x}$, $\hat{\sigma}_{j}^{y}$, and $\hat{\sigma}_{j}^{z}$ are Pauli matrices acting on the $j$th site. The longitudinal field is random and complex, $\mathcal{D}_j=\Delta_j + {\rm i} \gamma_j$, where the real part $\Delta_j$ and imaginary part $\gamma_j$ give the detuning and the gain or loss, respectively. The parameters $\Delta_j$ and $\gamma_j$ are independently and uniformly distributed within $[-W_\Delta,W_\Delta]$ and $[-W_\gamma,W_\gamma]$, correspondingly. In the following calculations, we scale the Hamiltonian with the XY interaction strength $J$. To explore the non-Hermitian effects, we vary $W_\gamma$ while fixing $h=0.5$ and $W_\Delta=1$ throughout the work. The model can be realized with Rydberg atom arrays, which will be discussed later. 

In the Krylov space, the original physical model is mapped to a single particle moving on a semi-infinite chain. Starting from the left end, the particle is hopping on the chain, giving the so-called Krylov wave function. As a result, the complexity growth is depicted by the propagation of Krylov wave function [see Fig.~\ref{fig:model}(b)]. The basis of Krylov space can be generated by various Krylov algorithms (see Appendix~\ref{appendix:Krylov} for details). The non-Hermitian Hamiltonian displays vastly different dynamics as the disorder strength $W_\gamma$ increases. Specifically, two phase transitions are observed, where a suppression of spreading within finite Krylov subspace takes place at $W_\gamma=W_L\approx 0.0351$ revealing a localization feature, and a transition from chaotic to nonchaotic dynamics at $W_\gamma=W_C\approx 1.647$. The phase diagram is depicted in Fig.~\ref{fig:model}(c). These phases can be captured by the spreading of the Krylov wave function, $\varphi_n$, illustrated in Fig.~\ref{fig:model}(d) (see Appendix~\ref{appendix:KrylovWF} for more examples). For $W_\gamma < W_L$, $\varphi_n$ is fully delocalized across the entire space at late times ($t=1000$). For $W_\gamma > W_L$, conversely, a weak breaking of ergodicity may be indicated by localization with exponentially small occupation for $n \gtrsim \mathcal{O}(L^2)$. In the nonchaotic regime, $W_\gamma < W_C$, $\varphi_n$ is further confined below $n\lesssim \mathcal{O}(L)$. We also observe prethermal behavior in the delocalized regime, where $\varphi_n$ is confined near $n\approx 0$ at mid-term times ($t=300$), contrasting with other phases. Krylov complexity and Lanczos coefficients capture critical behaviors around the phase transition points. This allows us to determine the critical points through a number of measures, which will be discussed in detail in the following. 

\begin{figure*}[ht]
    \centering
    \includegraphics[width=0.96\linewidth]{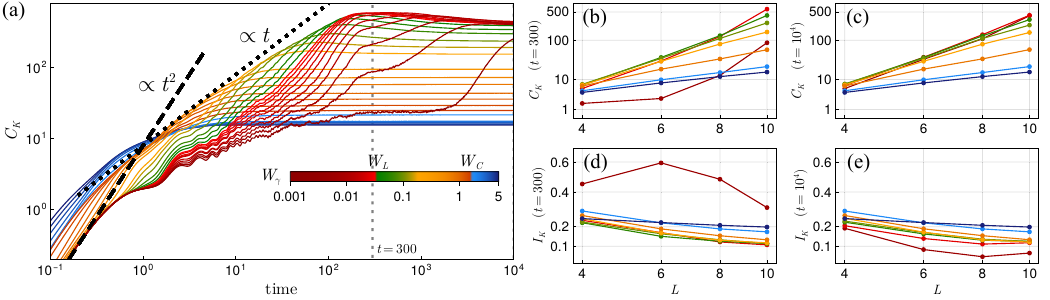}
    \caption{
    (a) Krylov complexity growth. $W_\gamma$ varies from $0.001$ to $5$ for colors from red to blue (from top to bottom in the long-time limit $t\sim 10^4$). We have fixed $W_\Delta=0.5$ and system size $L=10$. The dashed line is the quadratic ($\sim t^2$) trend and dotted line is the linear ($\sim t$) trend. The scaling of the Krylov complexity at (b) $t=300$ and (c) $t=10^4$ for different system sizes, and the colors, from red to blue, indicates the values of $W_\gamma = ( 0.005, 0.026, 0.06, 0.1, 0.2, 0.8, 1.8, 4.8 )$, in accordance with panel (a). (d), (e) The Krylov inverse participation ratios at the same times as panels (b) and (c).
    }
    \label{fig:complexity_growth}
\end{figure*}

\begin{table*}
  \begin{tabular}{c|c|c|c|c}
      \hline
       & Early-time growth  & Intermediate growth & Saturated value & Krylov localization\\ \hline 
      $W_\gamma<W_L$ & Quadratic & Plateau $\to$ Abrupt increase & High & Very high $\to$ low \\
      $W_L<W_\gamma<W_C$ & Quadratic & Linear envelope & Transitional & High \\
      $W_\gamma>W_C$ & Quadratic & Rapid saturation & Low & Very high \\
      \hline
  \end{tabular}
  \caption{\label{tb}
  A summary of major features of the complexity growth in different regimes of $W_\gamma$.
  }
\end{table*}

\section{Krylov complexity}
\label{sec:Krylov}

The mathematical framework for representing quantum models in Krylov space involves a tridiagonalization process~\cite{Liesen2013}. Krylov space is constructed by iteratively applying the Hamiltonian to an initial state vector and orthogonalizing it against the preceding basis vectors. This process is efficiently implemented using the bi-Lanczos algorithm, which is capable of dealing with non-Hermitian systems~\cite{Bhattacharya2023, Bhattacharjee2024, Carolan2024, Srivatsa2024}. In Krylov space, the evolution of the quantum system is encapsulated as an effective single-particle tight-binding model on a semi-infinite chain~\cite{Parker2019}. The wave function $\varphi_n$  is governed by the discrete equation (see Appendix~\ref{appendix:Krylov} for details),
\begin{equation}
    {\rm i} \frac{\partial}{\partial t} \varphi_n = b_{n} \varphi_{n-1} + a_n \varphi_n + c_{n+1} \varphi_{n+1},
\end{equation}
where $\varphi_n$ represents the wave function of the Krylov space. The Lanczos coefficients $b_n$ and $c_n$ are generally not complex conjugates, due to the non-Hermiticity of the Hamiltonian. The initial values of the Krylov wave function amplitudes are $\varphi_0=1$ and $\varphi_n =0$ for $n>0$.

Krylov complexity quantifies the spreading of $\varphi_n$ by its \emph{center of mass}, defined as
\begin{equation}
    C_K = \sum_n n \left|\varphi_n(t)\right|^2.
\end{equation}
We note that the norm of the wave function is not naturally conserved, but we always normalize the Krylov wave function $\varphi_n$ before calculating the Krylov complexity.
The growth of complexity is a rich, multistage phenomenon. Figure~\ref{fig:complexity_growth}(a) illustrates the complexity growth for system sizes $L=10$ with the initial state $\left|\psi_0\right\rangle= \left|+\right\rangle^{\otimes L}$, where $ \left|+\right\rangle = ( \left|\uparrow\right\rangle+ \left|\downarrow\right\rangle)/\sqrt{2}$. More examples of the complexity growth for different system sizes are shown in Appendix~\ref{appendix:KrylovWF}.
In case of Hermitian systems ($W_\gamma=0$), the complexity typically shows generic quadratic-linear-peak-saturate shape, reminiscent of the slope-dip-ramp-plateau profile of the spectral form factor~\cite{Erdmenger2023}. In our non-Hermitian spin chain model, instead of these trends, we find distinctive features, including Krylov localization, and chaos-nonchaos transition as $W_\gamma$ increases. In order to quantitatively depict the localization of the Krylov wave function, we employ the inverse participation ratio (IPR), defined as
\begin{equation}
    I_K = \sum_n \left|\varphi_n(t)\right|^4.
\end{equation}
The decreasing of $I_K$ indicates delocalization of $\varphi_n$, and its values at $t=300$ and $t=1000$ are shown in Figs.~\ref{fig:complexity_growth}(d) and \ref{fig:complexity_growth}(e).

\begin{figure}
    \centering
    \includegraphics[width=\linewidth]{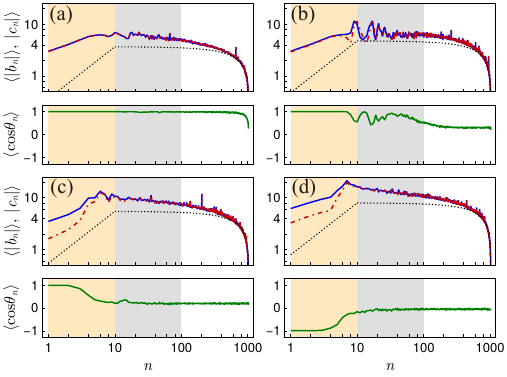}
    \caption{\label{fig:LanczosCoeffs}
    Lanczos coefficients for $L=10$ and $W_\gamma = $ (a) 0.0118, (b) 0.2, (c) 1.0, and (d) 3.0. Upper panels show $\langle |b_n|\rangle$ (solid blue) and $\langle |c_n|\rangle$ (dashed red), with dotted lines representing $\sim n$ for $n<L$, and $\sim \sqrt{1-n/2^L}$ for $n>L$. Lower panels show  $\langle \cos\theta_n\rangle$.
    }
\end{figure}

We identify three distinct regimes for the dissipation strength $W_\gamma$, separated by two critical values $W_L$ and $W_C$. The salient features of the complexity growth $C_K(t)$ and the localization of the Krylov wave function, characterized by the IPR, are contrasted in Table~\ref{tb}. Universally across all regimes, the short-time growth of complexity is quadratic, $C_K(t) \sim (1 + W_\gamma^2) t^2$ [dashed line in Fig.~\ref{fig:complexity_growth}(a)]; fitting details are provided in Appendix~\ref{appendix:early-time}. The absence of an exponential growth indicates a departure from the operator growth hypothesis. Combined with the nonlinear Lanczos coefficients shown in Fig.~\ref{fig:LanczosCoeffs}, this precludes a chaos diagnosis from early-time dynamics in our model. We therefore focus on the intermediate- and long-time behavior.

In the weak dissipative regime ($W_\gamma < W_L$), $C_K$ develops a pronounced plateau at intermediate times, whose height rises and duration shrinks as $W_\gamma$ increases [red curves in Fig.~\ref{fig:complexity_growth}(a)]. This signifies a prethermal stage where the Krylov wave function $\varphi_n$ is tightly confined at early times [Fig.~\ref{fig:model}(d), left], leading to an extraordinarily high IPR at $t=300$ [Fig.~\ref{fig:complexity_growth}(c)] that decreases only weakly with system size. Chaotic dynamics are not firmly established due to this mid-term plateau, but are expected to emerge in the long-time regime after $C_K$ abruptly increases to a high value, which is nearly independent of $W_\gamma$. Concurrently, $I_K$ decreases as the Krylov wave function finally spreads across the entire Krylov space. Thus, diagnosing chaos in the weak dissipative regime requires observing the long-time evolution of the complexity.

For moderate dissipation ($W_L < W_\gamma < W_C$), the prethermal plateau disappears. The intermediate growth of $C_K$ accelerates with increasing $W_\gamma$, yet its ultimate saturation value diminishes [green to yellow lines in Fig.~\ref{fig:complexity_growth}(a)]. This behavior forms a linear envelope, indicated by the dotted line in the same figure. In contrast to Ref.~\cite{Erdmenger2023}, where a linear ramp in complexity is linked to spectral rigidity and signifies quantum chaos, our results reflect chaos through a different form of complexity growth. This difference is also mirrored in the Lanczos coefficients. A linear ramp of $C_K$ results from $b_n \propto \sqrt{1-n/2^L}$, a special case of Tsallis $q$-log statistics with $q=0$~\cite{Bhattacharjee2024, Nandy2025}. Our model, however, yields $q>0$, as shown in Fig.~\ref{fig:LanczosCoeffs} and Appendix~\ref{appendix:Tsallis}. We also note that the suppression of the saturated complexity, the enhanced IPR, and the confinement of the Krylov wave function to $n < O(L^2)$ [Fig.~\ref{fig:model}(d), middle] suggest a weak breaking of ergodicity. This makes the chaotic dynamics for $W_L < W_\gamma < W_C$ distinct from those in the weak dissipation regime. The saturated $C_K$ exhibits a transitional scaling with system size [Fig.~\ref{fig:complexity_growth}(c)], the timescale to reach saturation is much shorter, and the chaotic evolution is restricted to a smaller Krylov subspace.

In the strongly dissipative regime ($W_\gamma > W_C$), $C_K$ saturates rapidly to a low value [blue lines in Fig.~\ref{fig:complexity_growth}(a)]. The Krylov wave function remains localized to $n < O(L)$ [Fig.~\ref{fig:model}(d), right], and the IPR is higher than that in the other regimes. These features signify a marked departure from conventional quantum chaos. As we will show later, the complex level statistics also confirm that $W_C$ marks the transition from chaotic to nonchaotic phases.

\section{Krylov variance and reciprocity}
\label{sec:results}

\begin{figure}
    \centering
    \includegraphics[width=\linewidth]{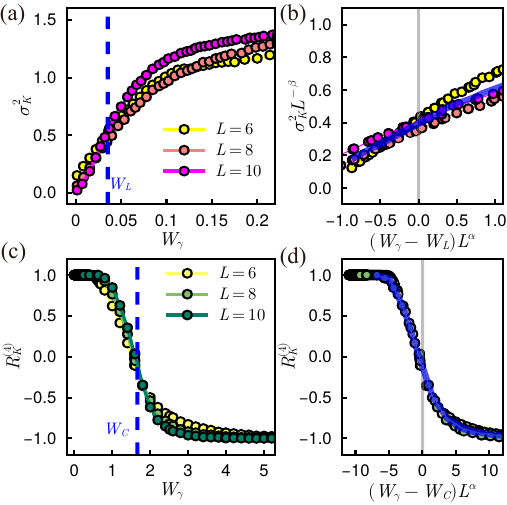}
    \caption{\label{fig:CriticalPoints}
        (a), (b) The localization to nonlocalization transition predicted by the Krylov variance $\sigma_K^2$, which exhibits a steep increase at the critical point, $W_L=0.0351$, determined by the finite-size scaling analysis (b).
        (c), (d) Chaotic-nonchaotic transition determined by the Krylov reciprocity $R_k^{(4)}$, which shows a sign flip at the critical point $W_C=1.647$. The result is consistent with the finite-size scaling (d).
    }
\end{figure}

In order to quantitatively determine the values of $W_L$ and $W_C$, we turn to the Lanczos coefficients.
Beyond the growth of $C_K$, a metric known as Krylov variance~\cite{Rabinovici2022} has been proposed to measure the fluctuation of the Lanczos coefficients, which characterizes the localization of the wave function as an analogue to Anderson localization~\cite{Fleishman1977}.
In Hermitian systems, the Krylov variance is defined as $\sigma_K^2= {\rm Var}\{\ln |b_{2n-1}/b_{2n}|\}$~\cite{Rabinovici2022}. Similarly, we define the variance of the non-Hermitian system by replacing $b_n$ with $j_n = |b_n c_n|^{1/2}$, yielding
\begin{equation} \label{eq:LanczosVariance}
    \sigma_K^2 = {\rm Var} \left\{ \ln \frac{j_{2n-1}}{j_{2n}} \right\} , \quad n = 1, 2, 3, \cdots .
\end{equation}
Figure~\ref{fig:CriticalPoints}(a) shows that $\sigma_K^2$ first grows with increasing $W_\gamma$ and saturates at larger $W_\gamma$.  Through finite-size scaling, we find that $\sigma_K^2$ can be universally fitted with function  $y=\sigma_K^2 L^{-\beta}$ against $x=(W_\gamma-W_L)L^\alpha$ [see Fig.~\ref{fig:CriticalPoints}(b)]. This allows us to identify a critical point, $W_L=0.0351$, with exponents $\alpha=1.816$ and $\beta=0.112$.
When $W_\gamma < W_L$, the variance is small, which results in the delocalization of the wave function, $\varphi_n$ [Fig.~\ref{fig:model}(d)]. When $W_\gamma > W_L$, $\varphi_n$ localizes in a finite region. Here, the system is still chaotic, as supported by the level repulsion (as discussed below). Yet, the localized wave function and hence the suppression of complexity emerge as a signature of ergodicity breaking~\cite{BallarTrigueros2022, Menzler2024, Li2025}.  

In non-Hermitian systems, Lanczos coefficients $b_n$ and $c_n$ are not complex conjugates, violating the reciprocity of the Krylov space. To quantify this, we evaluate arguments of the Lanczos coefficients, $\theta_n = \arg(b_n c_n)$, such that $\cos\theta_n = 1$ represents reciprocal tunneling, and $\cos\theta_n = -1$ gives the maximally nonreciprocal tunneling, which is associated with rapid relaxation. 
Figure~\ref{fig:LanczosCoeffs} shows the profiles of $\cos\theta_n$ as $W_\gamma$ increases.
Specifically, for $W_\gamma < W_L$ [Fig.~\ref{fig:LanczosCoeffs}(a)], the entire Krylov space remains reciprocal as $\braket{\cos\theta_n} \approx 1$ for all $n$, supporting complete delocalization of $\varphi_n$.
For $W_L < W_\gamma < W_C$ [Figs.~\ref{fig:LanczosCoeffs}(b) and \ref{fig:LanczosCoeffs}(c)], tunneling within $n<L$ is reciprocal as $\braket{\cos\theta_n} \approx 1$, but a rapid relaxation occurs as $\varphi_n$ spreads to $n>L$ with $\braket{\cos\theta_n} \approx 0.5$. 
For $W_\gamma > W_C$ [Fig.~\ref{fig:LanczosCoeffs}(d)], $\varphi_n$ rapidly relaxes within $n < L$, as the first few sites become maximally nonreciprocal with $\braket{\cos\theta_n} \approx 0$. This significantly inhibits the spreading of Krylov wave function and the complexity growth, much more influential than the fluctuations of the Lanczos coefficients.

To leverage the sign-flip behavior of $\cos\theta_n$, we introduce a parametrized metric,
\begin{equation} \label{eq:Reciprocity}
    R_K^{(d)} = \frac{1}{d} \sum_{n=1}^{d} \cos \theta_n  ,
\end{equation}
which is insensitive to the choice of $d$ for a certain range smaller than $L$. 
We show this reciprocal metric in Fig.~\ref{fig:CriticalPoints}(c) for $d=4$, and more examples in Appendix~\ref{appendix:Krylov_reciprocity}. For different $L$, $R_K^{(d)}$ collapses to $y = R_K^{(d)}$ with $x=(W_\gamma-W_C)L^\alpha$ [Fig.~\ref{fig:CriticalPoints}(d)], yielding the mean critical point $W_C=1.647$ and exponent $\alpha=0.820$ for $d\in\{4,5,6\}$, with the relative deviation $<1\%$. 
The proposed metrics from scaling analysis can signify the phase transitions linked to quantum chaos, and the phase diagram involving the two critical points, $W_L$ and $W_C$, is qualitatively illustrated in Fig.~\ref{fig:model}(c).

\section{Complex level spacing and entanglement entropy}
\label{sec:benchmark}

\begin{figure}
    \centering
    \includegraphics[width=\linewidth]{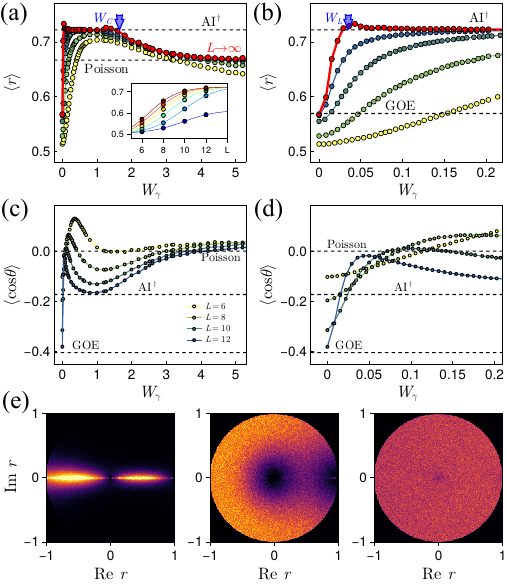}
    \caption{
    (a), (b) Mean radial distribution of complex level spacing ratio (CSR), $\langle r \rangle$, for different ranges of $W_\gamma$. Dot lines (red) are the extrapolated values when $L\to \infty$. See examples shown in the inset of panel (a), and details in Ref.~\cite{Note1}. Blue arrows indicate the phase transition points obtained from Krylov reciprocity for $W_C$ and Krylov variance for $W_L$, respectively.
    (c), (d) Mean angular distribution of CSR, $\langle \cos\theta \rangle$, for different ranges of $W_\gamma$. 
    (e) Distributions of complex level spacing ratios for $W_\gamma= 0.005,\ 0.8,$ and $\ 5.0$ and $L=12$ (from left to right), corresponding to GOE, AI${}^\dagger$ class, and 2D Poisson ensemble, respectively.
    }
    \label{fig:CSR}
\end{figure}

To provide a comprehensive view of the aforementioned phase transitions, we bridge the findings from the Krylov complexity with conventional signatures of quantum chaos including complex level spacing and entanglement entropy.

It is widely recognized to determine quantum chaos by the complex level spacing ratio (CSR), $r_j = (z_{j,{\rm NN}} - z_j)/(z_{j,{\rm NNN}}- z_j)$, where  $z_j$ is an eigenvalue of Hamiltonian~(\ref{eq:Hamiltonian}), and $z_{j,{\rm NN}}$ and $z_{j,{\rm NNN}}$ are its nearest-neighbor and next-nearest-neighbor eigenvalues~\cite{Sa2020}. The signatures of CSR are represented by its mean radial distribution $\langle r \rangle = \int |r|P(r){\rm d}r$ as shown in Figs.~\ref{fig:CSR}(a) and \ref{fig:CSR}(b), and the angular distribution by $\langle \cos\theta \rangle = \int \cos\theta P(r){\rm d}r$ ($\theta = \arg r$) shown in Figs.~\ref{fig:CSR}(c) and \ref{fig:CSR}(d). Increasing $W_\gamma$, the level statistics change from the Gaussian orthogonal ensemble (GOE) to ${\rm AI}^\dagger$ class, and then to the 2D Poisson ensemble [Fig.~\ref{fig:CSR}(e)]. To minimize the finite-size effects, we extrapolate $\langle r \rangle_\infty$ in the limit $L\to\infty$ [red dots in Figs.~\ref{fig:CSR}(a) and \ref{fig:CSR}(b)]
\bibnote{We assume the signatures of CSR for different system sizes read $\langle r \rangle = \langle r \rangle_\infty + \Delta{r} \left(1 + \exp[-a(L-L_0)] \right)^{-1}$ where $\langle r \rangle_\infty$, $\Delta{r}$, $a$, $L_0$ are fitting parameters. The values of $W_\gamma$ for the data in the inset are 0.01, 0.036, 0.064, 0.095, 0.126, 0.158 (from blue to red).}.
The values of $\langle \cos\theta \rangle$ do not show very clear scaling relations as $\langle r \rangle$ does, but also exhibit similar behavior to $\langle r \rangle$ as it transits between different values for different phases, and qualitatively reflects the spreading of the eigenvalues from the real axis to the whole complex plane.

As $W_\gamma \to 0$, $\langle r \rangle_\infty = 0.567$, consistent with $\langle r \rangle_{\rm GOE} = 0.5689$~\cite{Srivastava2019}. The angular signature is also close to $\langle \cos\theta \rangle_{\rm GOE} = -0.4038(2)$
\bibnote{To obtain the radial signature, we diagonalized $10^4$ symmetric random matrices of size $10^3 \times 10^3$. Using real matrices, we obtain $\langle \cos\theta \rangle_{\rm GOE} = -0.4038(2)$, and using complex matrices, we obtain $\langle \cos\theta \rangle_{\rm AI^\dagger} = -0.1727(2)$.}.
A plateau appears when $0.025 \lesssim W_\gamma \lesssim 1.6$, where $\langle r \rangle_\infty = 0.726(3)$, close to $\left\langle r \right\rangle_{\rm AI^\dagger} = 0.7222$~\cite{Garcia-Garcia2022}, and the angular signature $\langle \cos\theta \rangle$ also touches $\langle \cos\theta \rangle_{\rm AI^\dagger} = -0.1727(2)$ for large $L$. This indicates that the system belongs to the ${\rm AI}^\dagger$ class. 
Increasing $W_\gamma$ greater than $W_C$, the level statistics deviate from the AI${}^\dagger$ class and become the 2D Poisson ensemble. In this region, $\langle r \rangle_\infty = 0.6673(4)$, close to the 2D Poisson ensemble where $\left\langle r \right\rangle_{\rm Poi(2D)} = 2/3$ and $\left\langle \cos\theta \right\rangle_{\rm Poi(2D)} = 0$~\cite{Sa2020}. Therefore, the statistical signatures of CSR are consistent with the results of Krylov complexity.

\begin{figure}
    \centering
    \includegraphics{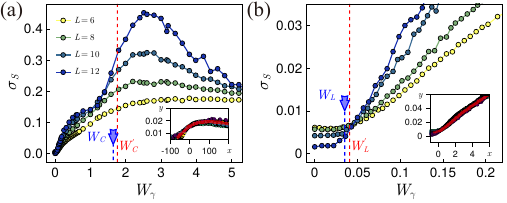}
    \caption{
        \label{fig:EE}
        (a) and (b) Standard deviation of bipartite entanglement entropy, $\sigma_S$, for different ranges of $W_\gamma$. The finite-size scaling [insets in (a) and (b)] yields critical points $W_C' = 1.763$ and $W_L' = 0.0409$. 
    }
\end{figure}

Another notable signature of phase transition is the bipartite entanglement entropy~\cite{Kjall2014}. We show the standard deviation of the bipartite entanglement entropy $\sigma_S$, in Fig.~\ref{fig:EE}(a). A divergent peak is observed as $L \to \infty$. The data can be universally described by function $y = \sigma_SL^{-\beta}$ with respect to $x = (W_\gamma-W'_C )L^\alpha$ [inset of Fig.~\ref{fig:EE}(a)]. This leads to the critical point $W'_C = 1.763$ with exponents $\alpha=2.013$ and $\beta=1.228$. 
Furthermore, a critical behavior at very small $W_\gamma$ can be seen in Fig.~\ref{fig:EE}(b). Using a similar scaling method, we identify the critical point $W'_L = 0.0409$ and exponents $\alpha=1.514$ and $\beta=-0.054$. The critical points extracted from $\sigma_S$ are similar to the ones of the Krylov complexity. In addition, we obtain a different critical point at $W^* = 3.236$ with the finite-size analysis of the entanglement entropy $S$ (see Appendix~\ref{appendix:entanglement_entropy}). This attributes to the transition to integrability as the system is well described by product states when $W_\gamma$ is sufficiently large.

\section{Conclusions}

In conclusion, we have studied non-Hermitian quantum chaos in the disordered XY model, revealing two critical phase transitions, including the chaotic-nonchaotic transition and Krylov localization. Through examining the Krylov complexity and Lanczos coefficients characterized by their variance and reciprocity, we identify the critical points and exponents of these phase transitions. Our results are corroborated by established measures including entanglement entropy and complex level spacings, which showcase the applicability of the complexity and associated measures in the study of quantum chaos. The sign flip of the reciprocity provides a quantitatively more precise criterion for chaos suppression than the entanglement entropy. Furthermore, we observe intricate chaotic behavior in non-Hermitian systems, where prethermalization may exist in the Krylov delocalized regime.

The non-Hermitian spin model can be realized in Rydberg atom arrays, where spin states are encoded by two Rydberg states~\cite{DeLeseleuc2019}. The XY interactions correspond to the dipole-dipole interactions between the Rydberg states, and the dissipation can be induced by coupling the Rydberg states to low-lying electronic states~\cite{Lourenco2022}. Randomness of the local fields can be introduced by ac Stark shifts with addressing lasers. In generic quantum simulators, non-Hermiticity can be realized through Sz.-Nagy dilation~\cite{Sweke2015, Hu2020a} and linear combination of unitaries~\cite{Childs2012, Schlimgen2021, Suri2023}. Finally, to measure the Krylov complexity, it has been proposed to apply measurable bases for estimation~\cite{Cindrak2024a}, and the measure of Loschmidt echo is potentially related to the ergodicity breaking [Fig.~\ref{fig:model}(d)].

\begin{acknowledgments} 
We are grateful to Dong-Ling Deng, Dong Yuan, Si Jiang, Chang Liu (SQZI), Chang Liu (NUS), Tianyi Yan, Xiaopeng Li, Xingze Qiu, and Saud \v{C}indrak for fruitful discussions.
Y.Z. acknowledges support from the Shanghai Qi Zhi Institute Innovation Program SQZ202317.
W.X. acknowledges support from the National Natural Science Foundation of China/Hong Kong RGC Collaborative Research Scheme (Project CRS CUHK401/22) and New Cornerstone Science Foundation. 
L.L. acknowledges support from the National Key Research and Development Program of China (Grant No. 2021YFA1402003) and the National Natural Science Foundation of China (Grants No. 12374329 and No. U21A6006).
W.L. acknowledges support from the EPSRC through Grant No.~EP/W015641/1, the Going Global Partnerships Programme of the British Council (Contract No.~IND/CONT/G/22-23/26), and the International Research Collaboration Fund of the University of Nottingham.
\end{acknowledgments} 


\appendix

\setcounter{figure}{0}
\renewcommand{\thefigure}{S\arabic{figure}}

\section{Krylov space and bi-Lanczos algorithm}
\label{appendix:Krylov}

\begin{algorithm}[b]
    \caption{Bi-Lanczos Algorithm with Complete Reorthogonalization}
    \label{alg:bi-Lanczos}
    \DontPrintSemicolon
    \SetAlgoLined
    \KwIn{$H$, $\psi_0$}
    \KwOut{Lanczos coefficients: $\{a_n\}$, $\{b_n\}$, $\{c_n\}$; Krylov basis: $\{p_n\}$, $\{q_n\}$}
    $b_0, c_0 \gets 0$, \quad $p_0, q_0 \gets \psi_0$ \tcp*{Initialize}
    \For(\tcp*[f]{$d=2^L$ is the dimension of the Hilbert space and $L$ is the system size}){$n = 0$ to $d-2$ }{
        $p_{n+1},\ q_{n+1} \gets H p_n,\  H^\dagger q_n$ \;
        $a_n \gets q_n^\dagger p_{n+1}$ \;
        $p_{n+1} \gets p_{n+1} - a_n p_n - c_{n} p_{n-1}$  \tcp*{$p_{-1} = 0$}
        $q_{n+1} \gets q_{n+1} - a_n^* q_n - b_{n}^* q_{n-1}$  \tcp*{$q_{-1} = 0$}
        $W \gets [p_0, p_1, \cdots, p_{n}] [q_0, q_1,\cdots, q_{n}]^\dagger $ \tcp*{Complete reorthogonalization}
        $\text{res} \gets 0$ \;
        \While{$\text{res} < 0.707$}{
            $\tilde{p}_{n+1} \gets p_{n+1} - W p_{n+1}$ \;
            $\tilde{q}_{n+1} \gets q_{n+1} - W^\dagger q_{n+1}$ \;
            $\text{res} \gets \min \left(\lVert\tilde{p}_{n+1}\rVert/\lVert p_{n+1} \rVert, \lVert\tilde{q}_{n+1}\rVert/\lVert q_{n+1} \rVert\right)$ \;
            $p_{n+1},\ q_{n+1} \gets \tilde{p}_{n+1},\ \tilde{q}_{n+1}$ \;
        }
        $b_{n+1} \gets \lVert p_{n+1} \rVert$ \tcp*{The argument of $b_n$ is fixed to be $0$.}
        $c_{n+1} \gets q_{n+1}^\dagger p_{n+1} / b_{n+1}$ \;
        $p_{n+1},\ q_{n+1} \gets b_{n+1}^{-1} p_{n+1},\ c_{n+1}^{*-1} q_{n+1}$ \;
    }
    $a_{d-1} \gets q_{d-1}^\dagger H p_{d-1}$ \;
\end{algorithm}

The study of Krylov complexity has its roots in solving Schrödinger equations, which starts with the temporal evolution of the wave function,
\begin{equation}
    \psi \left( t \right)  = e^{-iHt} \psi_0  = \sum_n{\frac{\left( -it \right) ^n}{n!} H^n \psi_0 } .
\end{equation}
In this context, the Hamiltonian $H$ and the initial wave function $\psi_0$ spans a Krylov space,
\begin{equation}
    \mathcal{K}(H,\psi_0) = \mathrm{span}\{\psi_0, H\psi_0, H^2\psi_0, \cdots \}.
\end{equation}
Contrasting with the Hermitian scenario, a non-Hermitian Hamiltonian generates a bi-orthogonal system, which includes an additional dual space,
\begin{equation}
    \mathcal{K}^*(H,\psi_0) = \mathrm{span}\{\psi_0, H^\dagger\psi_0, H^\dagger{}^2\psi_0, \cdots \}.
\end{equation}
These spaces can be orthogonalized to yield two sets of vectors, $\mathrm{span}\{p_0, p_1, p_2, \cdots\}$ and $\mathrm{span}\{q_0, q_1, q_2, \cdots\}$, which fulfill the bi-orthogonality condition,
\begin{equation} \label{eq:qp}
    q_m^\dagger p_n = \delta_{mn}.
\end{equation}
It is noteworthy that the corresponding matrices $P=[p_0,p_1,p_2,\cdots]$ and $Q=[q_0,q_1,q_2,\cdots]$ are not unitary. Consequently, the Hamiltonian can be tridiagonalized through the relation,
\begin{equation} \label{eq:tri}
    T = Q^\dagger H P,
\end{equation}
with $Q^\dagger P = P Q^\dagger = I$ given that $P$ and $Q$ are both square and invertible.

This process can be numerically executed via the bi-Lanczos (two-sided Lanczos) iteration. Here, we summarize the basic idea of bi-Lanczos algorithm. First, we assume that we have finished $n$ iterations, and obtained biorthonormal basis $\mathcal{K}_n = \{p_0, p_1, \cdots, p_{n}\}$ and $\mathcal{K}_n^* = \{q_0, q_1, \cdots, q_{n}\}$. Apart from Eq.~\eqref{eq:qp}, we can further let
\begin{equation}
    q_m^\dagger q_n = \delta_{mn} , \quad p_m^\dagger p_n \neq \delta_{mn},
\end{equation}
because only 2 out of 3 orthogonality conditions can be simultaneously satisfied in bilinear systems. 

Then, we can find the next basis $p_{n+1}$ by hitting $p_n$ with $H$, and removing its projection remained in $\mathcal{K}_{n}$. This procedure is similar to Gram-Schmidt elimination process. The projector is constructed by the bases in the dual space $\mathcal{K}^*_n$, and finally $H p_n$ can be written as
\begin{align}
    H p_n &= b_{n+1} p_{n+1} + \left( \sum_{m=0}^n q_m q_m^\dagger \right) H p_n, \nonumber \\
    &= b_{n+1} p_{n+1} + \sum_{m=0}^{n} \left( q_m^\dagger H p_n \right) q_m,
\end{align}
where $b_{n+1}$ is a normalization factor to be determined. Now, we consider the summation part. If $m < n-1$, $ q_m^\dagger H p_n = \left( H^\dagger q_m \right)^\dagger p_n = 0 $,
because $H^\dagger q_m \in \mathcal{K}^*_{m+1}$, $p_n \in \mathcal{K}_{n}$ but $p_n \notin \mathcal{K}_{m+1}$. Therefore, the summation over $m$ can be reduced to $m=n-1$ and $n$ only, and we eventually have a three-term recurrence relation 
\begin{equation}
    H p_n = b_{n+1} p_{n+1} + a_n p_n + c_{n} p_{n-1},
\end{equation}
with $a_n = q_n^\dagger H p_n$ and $c_n = q_{n-1}^\dagger H p_n$. The coefficients $b_n$, $a_n$, and $c_n$ are the diagonal and sub-diagonal elements of the tridiagonal matrix $T$ in Eq.~\eqref{eq:tri}. This recurrence relation can also be represented as $HP = PT$, and with the help of Eq.~\eqref{eq:tri}, we can also obtain the recurrence relation for $q_n$ which reads
\begin{equation}
    H^\dagger q_n = c^*_{n+1} q_{n+1} + a^*_n p_n + b^*_{n} q_{n-1}.
\end{equation}
To circumvent the breakdown issue, we have employed the complete reorthogonalization algorithm in this study. The full pseudo-code of this algorithm is shown in Algorithm~\ref{alg:bi-Lanczos}. We note that there are also several more expeditious methods, such as selective and partial reorthogonalization~\cite{VanDerVeen1995, Larsen1998, Liesen2013}.

With this methodology, the amplitude of the wave function within the Krylov space is defined by,
\begin{equation}
    \varphi_n(t) = q_n^\dagger \psi(t) .
\end{equation}
The temporal evolution of the coefficients $\varphi_n(t)$, akin to an effective tight-binding model, reads
\begin{equation}
    i \frac{\partial}{\partial t}\varphi_n = b_{n} \varphi_{n-1} + a_n \varphi_n + c_{n+1} \varphi_{n+1}.
\end{equation}

We note that the Arnoldi iteration, also known as the modified Gram-Schmidt algorithm, can deal with more general forms of $H$ with good numerical stability, albeit resulting in a more intricate recurrence relation, where the tridiagonal matrix $T$ is replaced by a Hessenberg matrix.

\section{Krylov wave functions}
\label{appendix:KrylovWF}

\begin{widetext}
    \begin{minipage}{\linewidth}
        \begin{figure}[H]
            \centering
            \includegraphics[width=\linewidth]{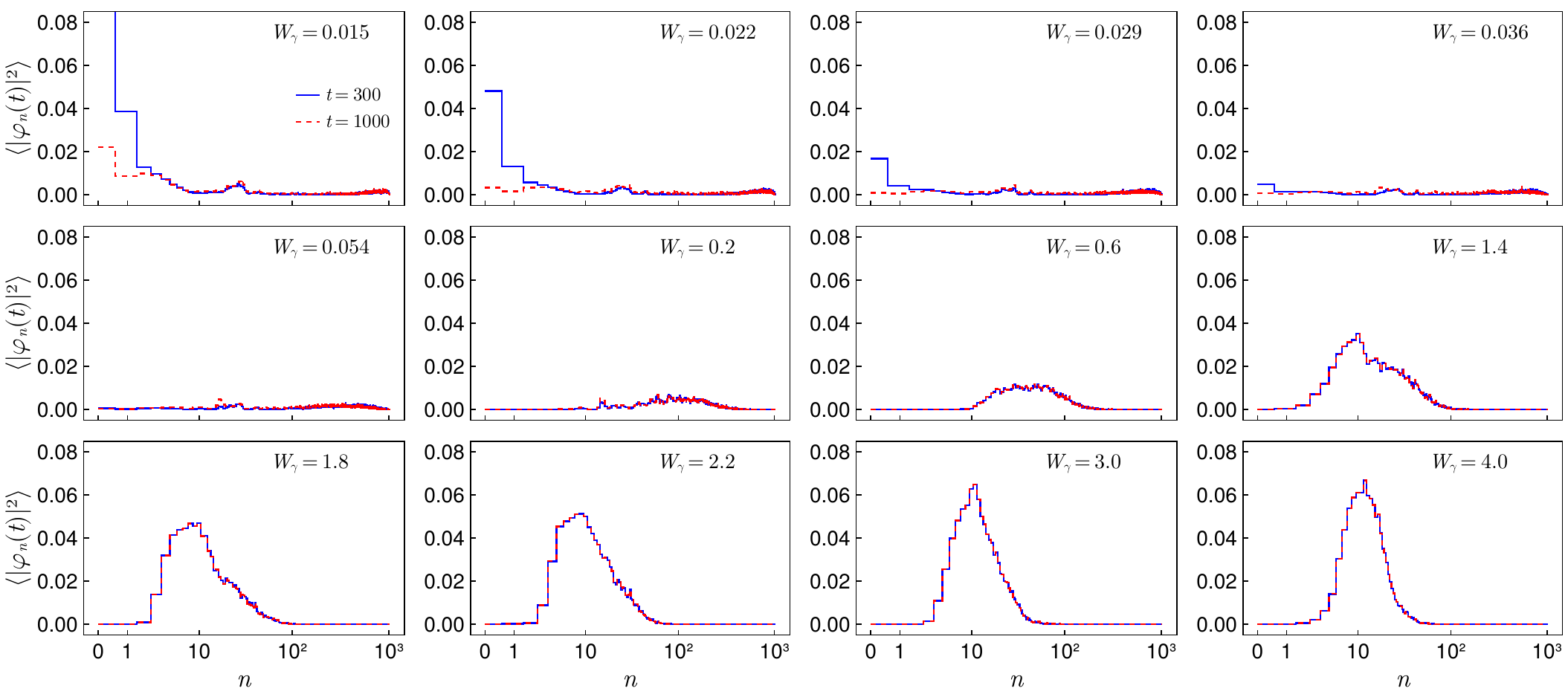}
            \caption{Krylov wave function at intermediate time $t=300$ (blue) and long time $t=1000$ (red) for different $W_\gamma$ and fixed $L=10$. }
            \label{fig:Krylov_WF}
        \end{figure}    
    \end{minipage}
\end{widetext}
    
Figure~\ref{fig:Krylov_WF} presents several examples of the Krylov wave function for varying $W_\gamma$ at times $t=300$ and $1000$, providing additional data depicted in Fig.~\ref{fig:model}(c). Our analysis here corroborates that the wave function exhibits localization around $n\approx 0$ for $W_\gamma < W_L$ ($W_L \approx 0.0351$) on the intermediate time scale ($t=300$), while delocalization is observed in the long term ($t=1000$). The apparent flatness of the delocalized wave functions is attributed to the logarithmic scaling of the $x$-axis, with significant occupation occurring at large $n$ values.

When $W_\gamma$ is comparable or larger than $W_L$, the Krylov wave function rapidly reaches a steady state characterized by localized features. The difference between the wave functions at $t=300$ and $t=1000$ are invisible. It is noted that the central peak shifts to smaller $n$ values as $W_\gamma$ increases.

When $W_\gamma > W_C$ ($W_C=1.647$), the saturated Krylov complexity remains relatively constant. The critical behavior may not be directly reflected in the Krylov wave function, however, as elaborated in the main text, it can be inferred from the Lanczos coefficients. We introduce the concept of Krylov reciprocity to identify the critical point, which provides a deeper understanding of the system at the phase transition.

\begin{figure}
    \centering
    \includegraphics[width=0.8\linewidth]{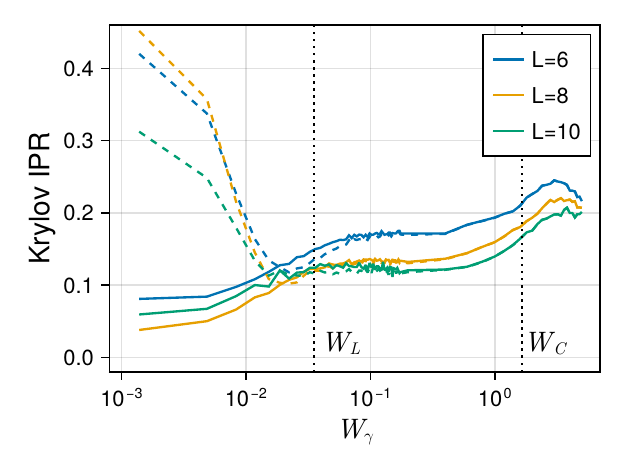}
    \caption{Krylov inverse participation at intermediate time $t=300$ (dashed) and long time $t=1000$ (solid) for different $W_\gamma$ and system size $L=6,8,10$. }
    \label{fig:Krylov_IPR}
\end{figure}

To quantitatively evaluate the localization of the Krylov wave function, we show the Krylov inverse participation ratio (IPR) in Fig.~\ref{fig:Krylov_IPR} for $t=300$ and $t=1000$, respectively. With higher values of Krylov IPR, the Krylov wave function is more localized. At $t=300$, the Krylov IPR is very large for $W_\gamma < 10^{-2}$, which corresponds to a prethermal behavior where the Krylov wave function is localized around $n\approx 0 $. At $t=1000$, the Krylov IPR monotonically increases as $W_\gamma$ increases, indicating that the Krylov wave function becomes more localized. There are several plateaus for Krylov IPR, and the ramping regimes correspond to the critical points of the phase transitions at $W_L$ and $W_C$.

\section{Krylov complexity growth for different system size and at early times}
\label{appendix:early-time}

\begin{figure}
    \centering
    \includegraphics[width=\linewidth]{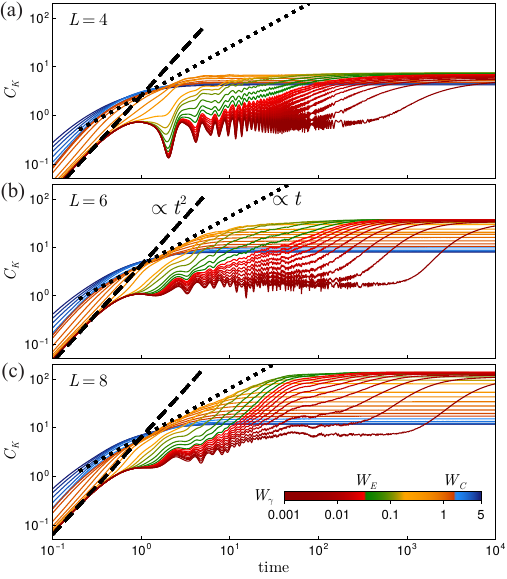}
    \caption{Growth of Krylov complexity for (a) $L=4$, (b) $L=6$ and (c) $L=8$. The dahsed lines are scaling of $\propto t^2$ and dashed-dot lines for $\propto t$.}
    \label{fig:CK_extended}
\end{figure}

Figure~\ref{fig:CK_extended} shows extended results for Krylov complexity growth across system sizes $L=4,\ 6,\ 8$, complementing the results in Fig.~\ref{fig:complexity_growth}(a). The observed dynamics closely match those presented in the main text. When $W_\gamma < W_L$, we consistently observe a plateau in $C_K(t)$ as $L$ increases, indicating prethermal behavior. A linear envelope applied to the mid-term growth for $W_L<W_\gamma<W_C$ across all system sizes. Furthermore, the long-time saturation values of $C_K(t)$ follow systematic scaling relationships with $L$, which are shown in Fig.~\ref{fig:complexity_growth}(b,c).

\begin{figure} 
    \centering
    \includegraphics[width=0.8\linewidth]{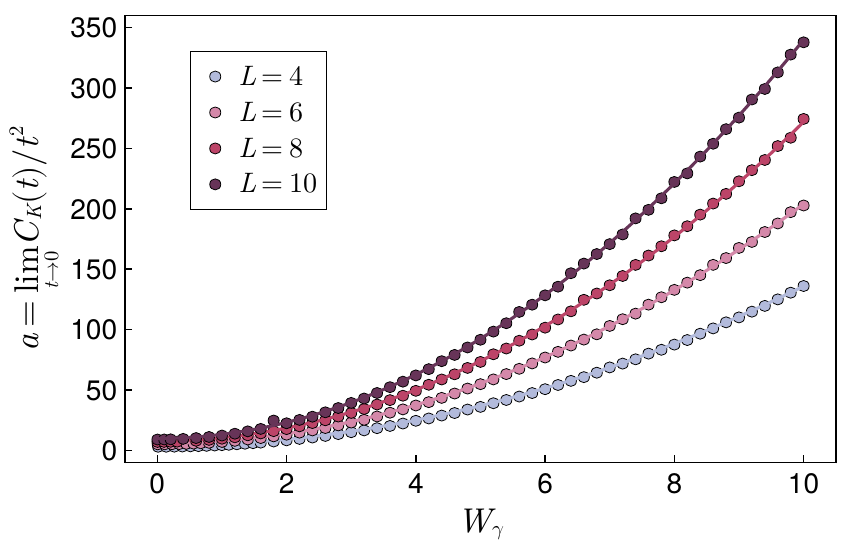}
    \caption{Early-time growth rate of the Krylov complexity, $a = \lim_{t\to 0} C_K / t^2$. Solid lines are the fitting with function $a = L (1 + \frac{1}{3} W_\gamma^2)$.  The fitting errors are smaller than sizes of the scatters.}
    \label{fig:early-time}
\end{figure}

The early-time growth of the Krylov complexity, $C_K$, exhibits quadratic behaviors. We fit $C_K$ using the function $C_K(t)=at^2$ for $t<0.01$, and present the coefficients for different system sizes $L=4,6,8,10$ and a range of $W_\gamma$ from 0.01 to 10 in Fig.~\ref{fig:early-time}. Our analysis reveals that the coefficient $a$ is proportional to $L$ and exhibits a parabolic relationship with $W_\gamma$, approximated as $a \simeq L (1 + 0.33  W_\gamma^2)$.

\section{Fitting of Lanczos coefficients with Tsallis $q$-log statistics}
\label{appendix:Tsallis}

\begin{figure}[b]
    \centering
    \includegraphics[width=\linewidth]{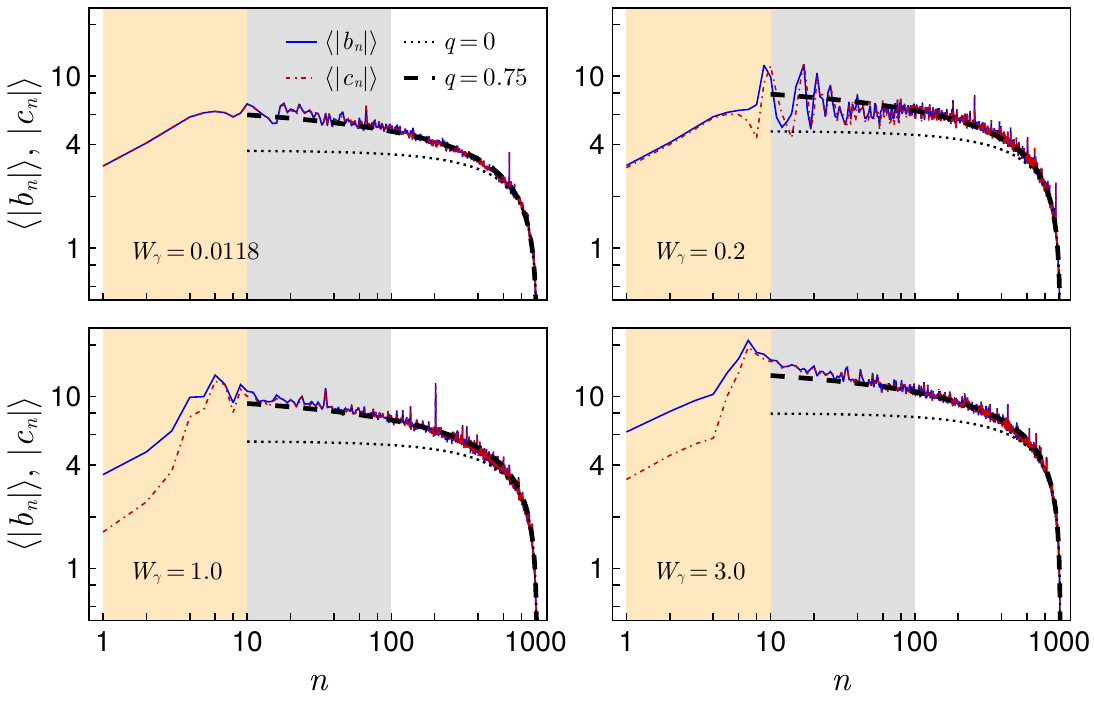}
    \caption{Fitting of the Lanczos coefficients with Tsallis $q$-log function. }
    \label{fig:Lanczos_fitting}
\end{figure}

Reference~\cite{Bhattacharjee2024} proposes using the Tsallis $q$-log function to fit the Lanczos coefficients for large $n$, with  $|b_n|^2 \propto -\ln_q(n/2^L)$. The $q$-log function is defined as
\begin{equation}
    \ln_q(x) = \begin{cases}\ln(x)&{\text{if }}x>0{\text{ and }}q=1\\{\frac {x^{1-q}-1}{1-q}}&{\text{if }}x>0{\text{ and }}q\neq 1\\{\text{Undefined }}&{\text{if }}x\leq 0\\\end{cases} .
\end{equation}
Thus, the assumption  $b_n \propto \sqrt{1-n/s^L}$ in Ref.~\cite{Erdmenger2023} can be regarded as a special case with $q=0$. We fit the average Lanczos coefficients $\langle |b_n| \rangle$ and  $\langle |c_n| \rangle$ using the form
\begin{equation}
\langle |b_n| \rangle, \langle |c_n| \rangle \propto \sqrt{1 - (n/2^L)^{1 - q}},
\end{equation}
with $q=0$ and $0.75$, as shown in Fig.~\ref{fig:Lanczos_fitting}. We find that $q=0.75$ provides a better approximation for the values of Lanczos coefficients in the regime where $n>L$, as discussed in the main text. 

\section{Krylov reciprocity}
\label{appendix:Krylov_reciprocity}

In order to show the $d$-dependence of the Krylov reciprocity, we show the values of $R_K^{(d)}$ for $d=3$ to $8$ in Fig.~\ref{fig:RK}(a1-a6). The crossing behavior is universal for all the choices of $d$, while their finite-size scaling has some small differences as shown in Fig.~\ref{fig:RK}(b,c).

\begin{widetext}
    \begin{minipage}{\linewidth}
        \begin{figure}[H]
            \centering
            \includegraphics[width=\linewidth]{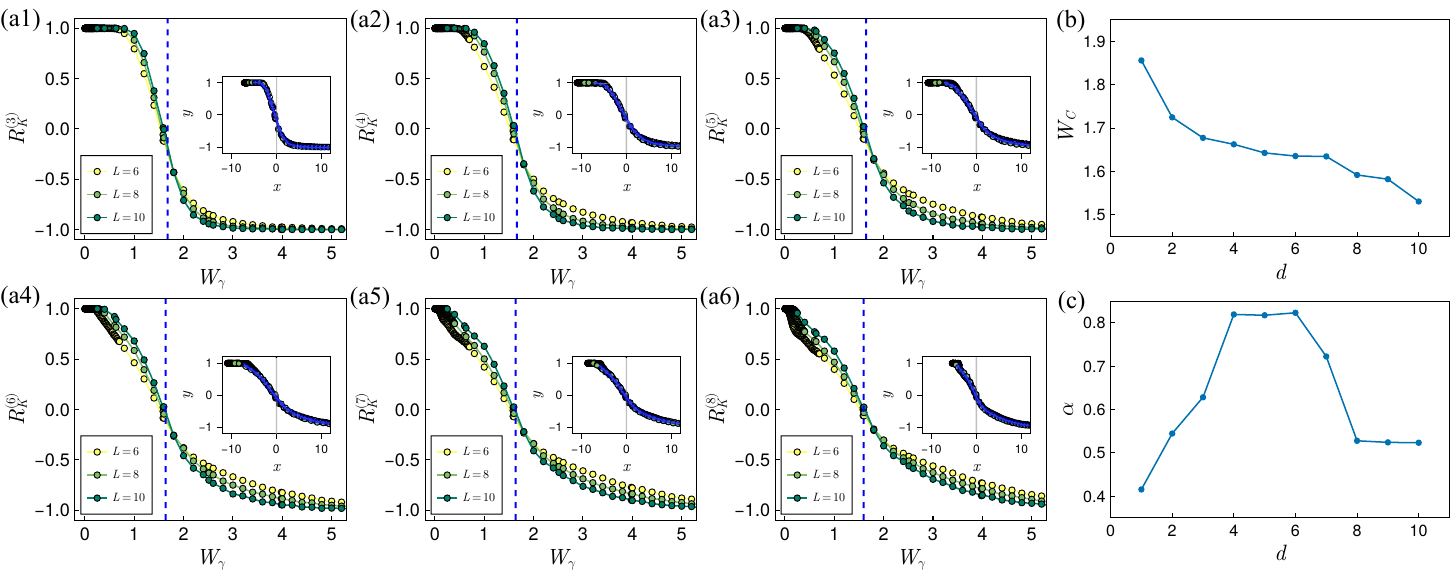}
            \caption{
            (a1-a6) Krylov reciprocity $R_K^{(d)}$ for $d=3$ to $8$. The insets are scaling collapse results for $x=(W_\gamma-W_C)L^\alpha$ and $y=R_K^{(d)}$. The critical points $W_C$ (b) and exponents $\alpha$ (c) are shown to have a shared plateau for $d=4$ to $6$, which are selected for identification of phase transition.
            }
            \label{fig:RK}
        \end{figure}    
    \end{minipage}
\end{widetext}

Here, we note that the finite-size collapse is carried out by minimizing the cosine similarity of the scaled data with the Nelder-Mead algorithm. This method is also applied to other finite-scaling analysis throughout this work.

We find that for $d=3$ to $7$, the collapsed critical point, $W_C$, exhibits a plateau, and there is a plateau for $d=4$ to $6$ for the critical exponent, $\alpha$. Taking the average values for $d=4$ to $6$, we find that $\langle W_C \rangle = 1.647 \pm 0.014$, and $\langle \alpha \rangle = 0.820 \pm 0.003$. The relative mean deviations are $0.85\%$ and $0.36\%$ for $\langle W_C \rangle $ and $\langle \alpha \rangle$, respectively.

\section{Bipartite entanglement entropy}
\label{appendix:entanglement_entropy}

In this section, we present the bipartite entanglement entropy of each eigenstate, as depicted in Fig.~\ref{fig:S_vs_E}. The entanglement is quantified using the von Neumann entropy for half of the chain, $S = - \textrm{Tr} \left( \rho_{L/2} \ln \rho_{L/2} \right)$,
where the reduced density matrix $\rho_{L/2}$ is obtained by tracing out the degrees of freedom of the chain's other half with indices $j>L/2$. The variance of entanglement entropy is taken by deviation of the mean entropy over different random configurations~\cite{Kjall2014}.

\begin{widetext}
    \begin{minipage}{\linewidth}
        \begin{figure}[H]
            \centering
            \includegraphics[width=\linewidth]{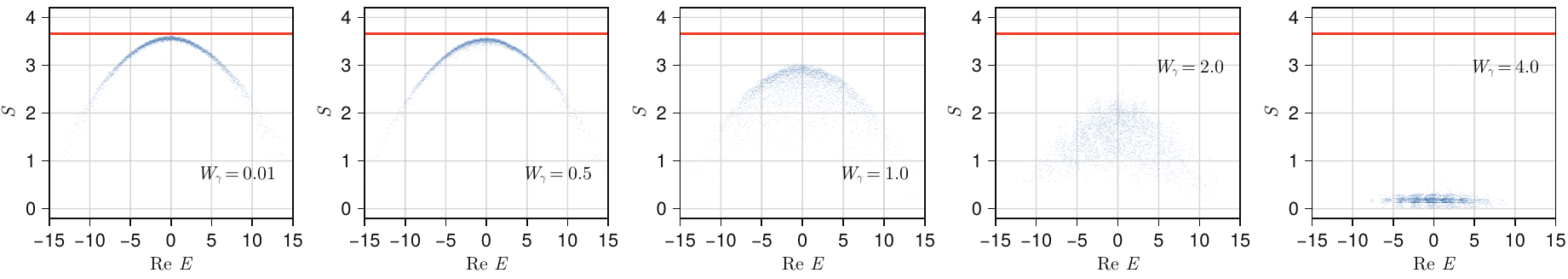}
            \caption{Bipartite entanglement entropy versus the real part of the eigenvalues of each eigenstate. The orange lines are the Page value. Here we set $L=12$ and $S_{\rm Page} = 6 \ln2 - 1/2 \approx 3.659$.}
            \label{fig:S_vs_E}
        \end{figure}    
    \end{minipage}
\end{widetext}

Generally, in the chaotic regime, the bipartite entanglement entropy scales extensively with subsystem volume, which is half of the total system size, $L/2$. The maximum entropy approaches the Page value, $S_{\rm Page} = (L \ln 2 - 1)/2$~\cite{Kjall2014}. As disorder strength $W_\gamma$ increases and the system transitions to the nonchaotic phase, the entanglement entropy of eigenstates decreases and their distribution narrows significantly. This compression shifts the shape of the distribution from a broad arc to a cigar-like profile. At extreme disorder (e.g., $W_\gamma=4.0$), the entropy approaches zero, signaling eigenstates that are nearly product states. This near-factorization of wavefunctions strongly suggests the emergence of integrability in the strongly dissipative limit.

\begin{figure}[b]
    \centering
    \includegraphics[width=\linewidth]{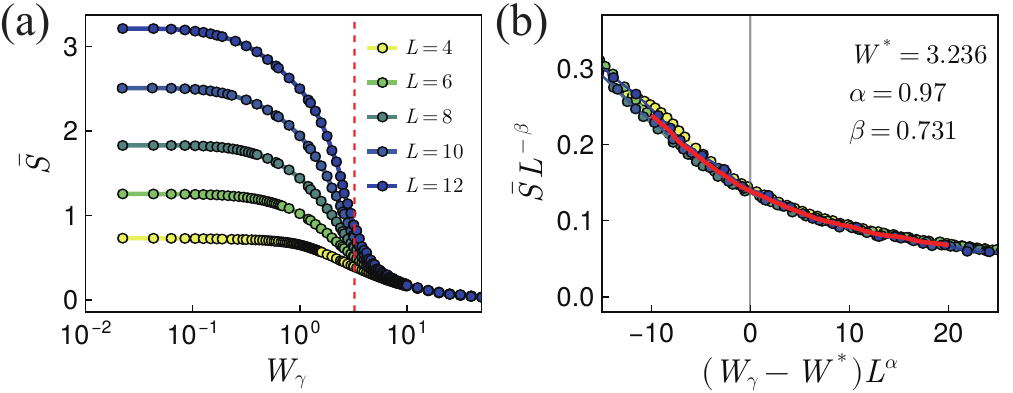}
    \caption{(a) Bipartite entanglement entropy for varying system size $L$ and non-Hermitian disorder $W_\gamma$. The red dashed line gives $W^*$ obtained from the finite-size scaling. (b) Finite-size scaling of $S$. The Red line is the collapsed universal function by averaging data within the interval for optimization.}
    \label{fig:entanglement_entropy}
\end{figure}

In the chaotic regimes where $W_\gamma$ is small, the mean value of the entanglement entropy, $\bar{S}$, increases as shown in Fig.~\ref{fig:entanglement_entropy}(a). At this point, no discernible difference is observed between Hermitian and non-Hermitian chaos, as $S$ approaches constant values when $W_\gamma\to0$. As $W_\gamma$ increases, $\bar{S}$ tends to diminish, which is consistent with the hypothesis that the system's wave function evolves into product states. Additionally, we have conducted a finite-size scaling analysis, as shown in Fig.~\ref{fig:entanglement_entropy}(b). The critical value $W^*$ as well as the exponents $\alpha$ and $\beta$ are obtained by finite-size scaling and $W^*=3.236$ is in proximity to the transition point where the characteristic complex level spacing $\langle r\rangle$ descends to $2/3$, indicative of a 2D Poisson distribution.

Conversely, the examination of the standard deviation of the bipartite entanglement entropy, $\sigma_S$, along with Krylov reciprocity, indicates a phase transition at $W_C \approx 1.763$ and $1.647$, respectively, as it departs from the chaotic regime of the AI$^\dagger$ class where $\langle r \rangle_{\textrm{AI}^\dagger} = 0.7222$.


%

\end{document}